\pdfoutput=1
\documentclass[10pt,conference,letterpaper]{IEEEtran}

\IEEEoverridecommandlockouts
\usepackage{cite}
\usepackage{hyperref}
\hypersetup{
colorlinks=true,
linkcolor=black,
filecolor=black,
urlcolor=black,
citecolor=black,
}
\usepackage{amsmath,amssymb,amsfonts}
\usepackage{algorithmic}
\usepackage{graphicx}
\usepackage{textcomp}
\usepackage{amssymb}
\usepackage{amsmath}
\let\subfigure\relax
\usepackage[ruled]{algorithm2e}
\usepackage{xparse}
\usepackage{makecell}
\usepackage{multirow}
\usepackage{balance}
\usepackage{mathtools}
\usepackage{subfigure}
\usepackage{graphicx}
\usepackage{textcomp}
\usepackage{xcolor}
\usepackage{float}
\usepackage{bm}
\usepackage{soul}
\usepackage{xspace}
\usepackage{comment}
\usepackage{bm}
\usepackage{ragged2e}
\usepackage{amsmath}
\usepackage{enumitem}

\usepackage{geometry}

\usepackage{threeparttable}
\usepackage{utfsym}

\newcommand{\hjy}[1]{{\color{black}#1}}

\def\BibTeX{{\rm B\kern-.05em{\sc i\kern-.025em b}\kern-.08em
    T\kern-.1667em\lower.7ex\hbox{E}\kern-.125emX}}

\ifCLASSOPTIONcompsoc
  \usepackage[nocompress]{cite}
\else
  \usepackage{cite}
\fi

\ifCLASSINFOpdf

\else

\fi

\newcommand{\name}{Tremerity-Fi\xspace}

\begin{document}

\title{\name: Non-Contact Daily-Life Tremor Severity Assessment by Commercial mmWave Radar \\

\author{
Xiao Li$^{1}$, Jingyang Hu$^{1}$, Shang Gao$^{1}$, Yichao Gao$^{1}$, Yiyu Xin$^{1}$, Hongbo Jiang$^{2}$, Xiang-Yang Li$^{1}$\\
$^{1}$School of Computer Science and Technology, University of Science and Technology of China, China\\
$^{2}$College of Computer Science and Electronic Engineering, Hunan University, China\\
Email: \{xiao\_li, hujingyang, shang\_gao, gyc77, xyy\_2002\}@mail.ustc.edu.cn,\\ hongbojiang2004@gmail.com, xiangyangli@ustc.edu.cn
}

}

\maketitle

\begin{abstract}
Tremor is a common symptom of neurological diseases. The regular assessment of daily tremors facilitates the evaluation of disease progression and assists clinicians in optimizing treatment strategies.
However, current home monitoring solutions have difficulty in dealing with user cooperation, privacy concerns, environmental interference, and system generalization, leading to feasibility concerns in activities of daily living (ADL).
To this end, we propose \name, a non-contact and privacy-friendly tremor severity assessment system based on mmWave radar.
To realize \name, we first design an adaptive beamforming algorithm to accurately identify useful but weak signals from numerous reflections captured in the environment.
Second, unlike primary reflections commonly used in mmWave sensing, we leverage multipath reflections that carry useful information about the target’s motion, even though they are generally considered harmful, to help reconstruct hand signals and improve sensing performance.
Furthermore, we propose an unsupervised domain adaptation algorithm to improve the ability to adapt to unseen environments and users.
We collect a diverse dataset of \hjy{5} patients and \hjy{25} healthy subjects in \hjy{3} scenarios, such as offices, homes, and hospitals.
Extensive experiments show that our system achieves \hjy{94.51\%} accuracy in tremor detection, about \hjy{5}\% higher than the SOTA mmWave radar method, and \hjy{89.13\%} in tremor severity assessment, demonstrating its sufficient potential as a tremor monitoring assistant for patients with neurological diseases.

\end{abstract}


\section{Introduction}
\label{SEC:Introduction}

Tremor~\cite{elble2017tremor} is a common symptom of neurological diseases, including physiological tremor (PT), essential tremor (ET), Parkinson's disease (PD), etc. About 75\% of PD patients~\cite{Tremor} (more than 10 million patients worldwide) experience tremor at some point in the course of the disease. Tremor directly affects daily activities and seriously affects the quality of life~\cite{okelberry2024updates} of patients. Currently, symptoms can be effectively suppressed mainly by adjusting the treatment strategy and controlling the use of drugs according to the severity of tremor~\cite{yan2024degeneration}. Therefore, daily home tremor monitoring can provide doctors with an effective reference for medication adjustment and can greatly reduce the economic costs incurred by frequent trips to the hospital for patients.

Existing mainstream tremor monitoring methods can be divided into three categories: wearable device-based monitoring methods~\cite{timmermans2025generalizable, paredes2024upper,maas2024patient}, vision-based monitoring methods~\cite{xu2025improving,rabano2025digital}, and wireless-based monitoring methods~\cite{mejdani2024radar,li2024pd,gillani2021unobtrusive}. Although these methods have broad prospects, there are still limitations that restrict their development. Wearable device-based monitoring methods show high accuracy, but discomfort caused by the long-term wearing of dedicated equipment and poor user compliance remain problems~\cite{paredes2024upper}. Vision-based monitoring methods may raise serious privacy concerns~\cite{zhao2025visual} for patients, and their performance will deteriorate in scenes with limited lighting. RF-based monitoring methods have become a research hotspot due to their non-contact and privacy-friendly characteristics. However, existing methods always test simulated tremors~\cite{gillani2023unobtrusive} or overlook testing in non-line-of-sight (NLOS) scenarios.

\begin{figure}[t]
\centerline{\includegraphics[width=0.5\textwidth]{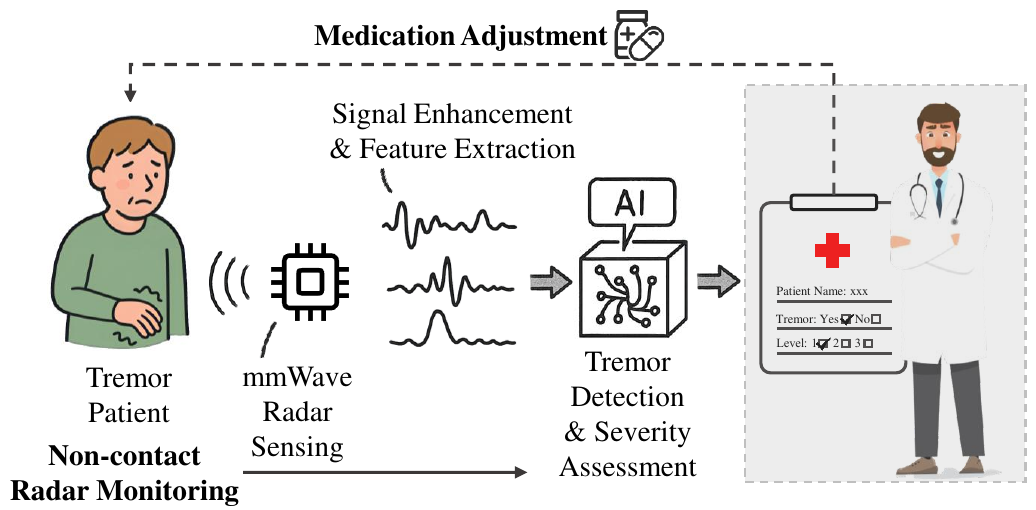}}
\vspace{-1em}
\caption{Usage of \name, which helps with medication by detecting tremors and assessing their severity using mmWave radar.}
\label{fig:TremorSense}
\vspace{-1em}
\end{figure}

In this work, we propose \name~(illustrated in Fig.~\ref{fig:TremorSense}), a novel non-contact, privacy-friendly tremor monitoring system based on mmWave radar, taking a step forward in promoting mmWave radar sensing technology to practical mobile health applications. When designing the \name system, we adhered to the following four principles:
(1) Non-contact: Ensures user comfort without disrupting daily activities, especially for PD patients with mobility issues;
(2) Privacy-friendly: Avoids capturing sensitive personal information, suitable for home monitoring;
(3) Accuracy: Maintains reliable performance under various interference conditions;
(4) Generalization: Delivers consistent accuracy across diverse users and environments.

Although \name is a promising system, several technical challenges still need to be overcome when it is actually applied to the home health monitoring scenario.

\begin{itemize}

    \item \textbf{Hand reflection signal extraction:~}Accurate tremor assessment relies on the reliable extraction of fine-grained motion from the subject's hand~\cite{elble2006tremor}. However, compared to the strong reflections from larger body parts, the hand's reflective surface is relatively small, making it challenging to accurately isolate the hand's reflected signals.

    \item \textbf{Signal distortion:~}In real-world environments, the subject’s hand is sometimes partially occluded by objects, which can compromise the performance of tremor monitoring. Enhancing hand tremor signals under such NLOS conditions presents a significant challenge.

    \item \textbf{System generalization:~}The collection and analysis of tremor data are greatly affected by inter-subject variability (e.g., tremor amplitude and frequency) and environmental factors (e.g., background noise and interference). These factors present substantial challenges to the generalizability of the data across different populations and scenarios.

\end{itemize}

To tackle the first challenge of isolating weak hand tremor signals, we analyze the spatial and frequency characteristics of mmWave radar reflections. We observe that dominant signals often originate from the chest, a large static reflector, while hand tremor signals are weaker. By applying clutter removal, we eliminate the interference from the static background. An adaptive beamforming algorithm, guided by the spatial prior that the hand lies in front of the chest, is used to localize the hand. Multi-angle signal aggregation is then applied to enhance the hand’s weak tremor reflections.

Second, to solve the signal distortion caused by occlusions, we propose a secondary reflection enhancement method. By leveraging multipath signals reflected off nearby surfaces, we identify and fuse tremor-relevant secondary paths based on spectral energy and phase similarity. This approach reconstructs a more stable hand signal, ensuring reliable tremor sensing even under partial blockage.

Third, to improve the generalization of the system, we adopt an adversarial domain adaptation framework. The model is trained to learn domain-invariant features that preserve tremor severity characteristics while remaining robust to variations in user identities, postures, and environments. This allows the system to perform accurately in unseen scenarios with minimal labeled data, enabling practical deployment.

We implement the prototype of \name using a commercial mmWave radar and laptop platform, and conduct a large number of experiments to evaluate its performance. Experimental results show that \name can achieve high-precision tremor presence detection and fine-grained tremor severity assessment. Overall, our main contributions include:

\begin{itemize}
    \item We introduce the first mmWave radar system for assessing tremor severity under various conditions, proven effective for real patients. This system uses only commercial mmWave radar and supports medication guidance.

    \item We develop a spatial prior-guided tremor source separation algorithm to isolate weak hand motion signals from environmental reflections and propose a secondary reflection enhancement method to improve the signal quality in NLOS conditions.

    \item We conduct extensive real-world experiments involving 30 subjects and 3 distinct environments. Experimental results demonstrate that \name~achieves an accuracy of 94.51\% in tremor detection and 89.13\% in tremor severity assessment, highlighting its robustness and effectiveness for practical deployment. Our experiments further show that the proposed secondary reflection enhancement method improves the system accuracy by approximately 15\% under NLOS conditions.

\end{itemize}

\section{Background and Feasibility Analysis}
\label{SEC:Feasibility}

\begin{figure*}[!t]
\centerline{\includegraphics[width=0.8\linewidth]{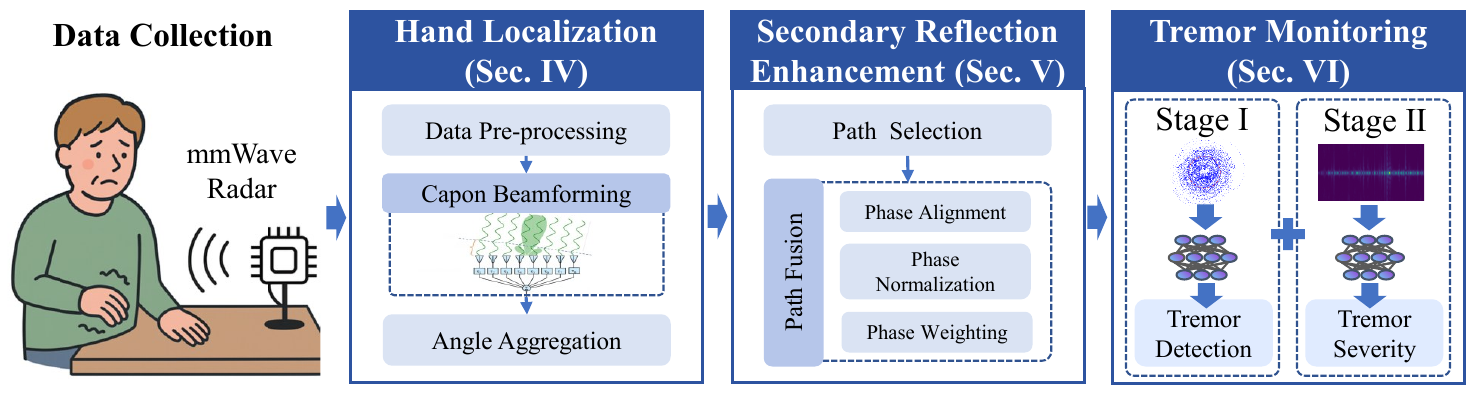}}
\vspace{-.8em}
\caption{Overview of \name. We first localize the hand (Sec.~\ref{SEC:Hand Localization}) and then enhance the primary signal using secondary reflections (Sec.~\ref{SEC:Secondary Relection Enhancement}). In tremor monitoring (Sec.~\ref{SEC:Tremor Monitoring}), IQ images are utilized for tremor detection, while STFT images are employed for tremor severity assessment.}
\label{fig:sytem}
\vspace{-1em}
\end{figure*}

\subsection{Clinical Background of Tremor Disorders}
Tremor is one of the most common motor symptoms in neurological disorders, often associated with PD and ET. While PD tremor typically occurs at rest and is low-frequency (3–6 Hz), ET tremor is characterized by higher frequency (4–12 Hz) postural or kinetic tremor. Clinically, tremor severity is graded using qualitative scales such as UPDRS (Unified Parkinson’s Disease Rating Scale)~\cite{martinez1994unified} or TETRAS (Tremor Research Group Essential Tremor Rating Assessment Scale)~\cite{ondo2020tremor}. Doctors adjust the medication of patients based on their tremor severity. Although these scales provide useful information, they are limited by their subjective nature and dependence on in-clinic assessments, which fail to capture the full extent of tremor severity during daily life.
Therefore, an accurate assessment of tremor severity in daily life is crucial for better management of patients with PD and ET.

\subsection{mmWave Radar Enabled Contactless Sensing}
MmWave radar can resolve submillimeter displacement by analyzing the phase variation of the backscattered signal. The round-trip phase shift induced by a target displacement $d(t)$ is $\phi(t) = \frac{4\pi d(t)}{\lambda}$, where $\lambda\!\approx\!5$\,mm at 60\,GHz. Thus, a displacement as small as 0.1\,mm yields a measurable phase shift of 0.25\,rad, enabling high-fidelity tracking of subtle tremor-induced motions. By computing the inter-chirp phase difference and performing phase unwrapping, fine-grained displacement and velocity trajectories can be reconstructed even within a single range bin, making mmWave radar well-suited for extracting oscillatory tremor waveforms. Moreover, its device-free and privacy-preserving nature makes it ideal for long-term deployment in homes, hospitals, and elder-care environments. Recent studies have demonstrated its broad potential in health monitoring, including respiration~\cite{zhang2022quantifying,liu2019beyond}, heart rate~\cite{zhang2022can,gong2025osense}, blood pressure~\cite{cao2024hbp,hu2024contactless}, and dry eye screening~\cite{xue2024sde}.

Motivated by the urgent need for an accurate tremor severity assessment method in daily life and the fine-grained motion sensing capability of mmWave radar, we investigate mmWave's application in tremor detection and severity assessment.

\section{System Overview}
\label{SEC:System Overview}

Fig.~\ref{fig:sytem} presents an overview of \name, which comprises three main components: \textit{Hand Localization}, \textit{Secondary Reflection Enhancement}, and \textit{Tremor Monitoring}.

\textit{Hand localization} aims to isolate the weak hand signal from interference caused by other body parts. First, we design a clutter removal algorithm to suppress reflections from static objects. Given the prior knowledge that the human chest acts as the dominant reflector\cite{adib2015capturing}, we apply the improved Capon beamforming to range bins preceding the chest to generate a high-resolution range-azimuth map, on which a clustering algorithm~\cite{deng2020dbscan} localizes the hand. Signals from multiple antenna channels are then coherently aggregated to further enhance the hand signal.

\textit{Secondary Reflection Enhancement} exploits secondary reflection paths to reinforce the primary reflection for tremor reconstruction, particularly when the line-of-sight (LoS) path is partially occluded. We first identify secondary components corresponding to the same target, align them in phase, and normalize to remove scale mismatches. A phase-weighted fusion strategy then combines the primary and secondary signals to improve signal quality and stability.

\textit{Tremor Monitoring} performs detection and severity assessment via a two-stage model. Stage~I takes the hand-region IQ image as input to detect tremor presence from motion patterns. If tremor is detected, Stage~II feeds the micro-Doppler spectrogram, obtained via Short-Time Fourier Transform (STFT), into a Domain Adversarial Neural Network (DANN)~\cite{ganin2016domain} to classify severity levels.
This division allows us to exploit the complementary strengths of time-domain and frequency-domain features for accurate and fine-grained tremor monitoring, while also reducing computational overhead by avoiding unnecessary severity classification when no tremor is present.

\section{Hand Localization}
\label{SEC:Hand Localization}
Hand signals are typically much weaker than reflections from other parts of the body, making accurate localization difficult. In this section, we present an advanced method designed to localize the hand that only uses mmWave radar.

\subsection{Data Pre-processing}
The received mmWave radar signals are first processed by FFT to separate the reflections from different distances. Clutter removal is then applied to suppress static components. To detect potential targets, we employ a Cell-Averaging Constant False Alarm Rate (CA-CFAR) algorithm with an adaptive threshold. Specifically, we introduce a percentile-based dynamic lower bound, defined as
\begin{equation}
    l_{\text{bound}} = \text{Percentile}(P, \alpha)
\end{equation}
where $P$ is the set of power values in the range profile, and $\alpha$ is a predefined percentile (e.g., 60 in our implementation). This dynamic threshold helps suppress residual noise and weak clutter more effectively than static thresholds.

\begin{figure}[!tb]
	\begin{tabular}{cc}
		\begin{minipage}[t]{0.45\linewidth}
			\includegraphics[width = 1\linewidth]{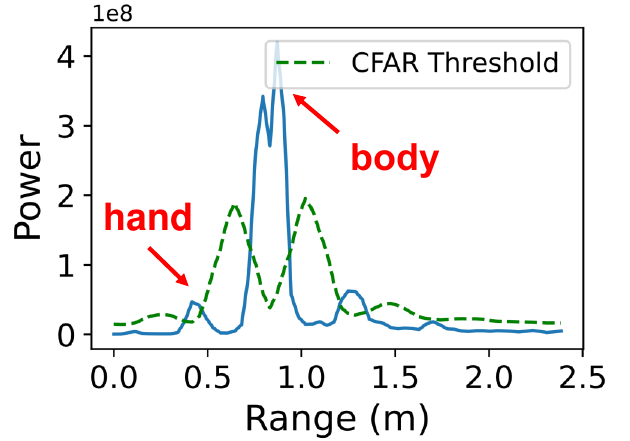}
                \vspace{-1.5em}
			\caption{Power bin map in the sensing range. Hand signal is much weaker than body signal.}
			\label{fig:power_bin}
		\end{minipage}
        \quad
		\begin{minipage}[t]{0.45\linewidth}
			\includegraphics[width = 1\linewidth]{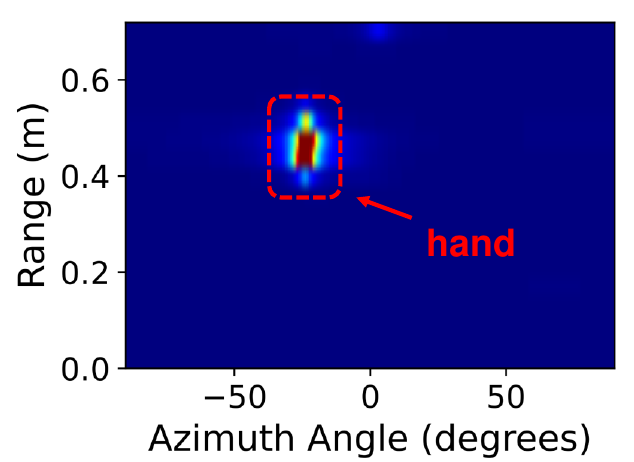}
                \vspace{-1.5em}
			\caption{Range-azimuth map of the hand after using the adaptive beamforming algorithm.}
			\label{fig:RA_map_hand}
		\end{minipage}
	\end{tabular}
    \vspace{-1em}
\end{figure}

\subsection{Capon Beamforming}
The human chest is the largest convex reflector on the human body~\cite{adib2015capturing}, typically generating the strongest return in the power bin map. As illustrated in Fig.~\ref{fig:power_bin}, hand reflections are much weaker than those from the chest. To extract the hand signal, we apply the Capon beamforming algorithm specifically to the range bins preceding the chest region, thereby generating a high-resolution range-azimuth map focused on the hand. This approach effectively suppresses interference from irrelevant areas and enhances the visibility of hand signals (see Fig.~\ref{fig:RA_map_hand}). To further refine the estimation of the hand’s angular position, we apply DBSCAN (Density-Based Spatial Clustering of Applications with Noise)~\cite{schubert2017dbscan} to cluster the beamformed points. The centroid of the identified cluster provides an accurate estimate of the hand's location.

\subsection{Multi-angle aggregation}
To further improve the quality and stability of weak hand tremor signals, we employ a multi-angle aggregation technique based on phase-coherent beamforming across receiving antennas. Given the estimated azimuth angle $\theta_{ri}$ of the hand and its corresponding range bin, we compute the phase offset between adjacent antennas as

\begin{equation}
    \Delta \phi_{ri} = \frac{2 \pi \lambda}{d_{int}} \sin(\theta_{ri})
\end{equation}
where $d_{int}$ denotes the distance between adjacent antennas. The signals from $N$ receiving antennas are then phase-aligned using the compensation term $e^{-j(n-1)\Delta \phi_{ri}}$ for antenna index $n$, and aggregated as

\begin{equation}
    S_{BF_{\theta_{ri}}}(t) = \sum_{n=1}^{N} S_{IF_n}(t) \exp\left( -j (n-1) \Delta \phi_{ri} \right)
\end{equation}

\begin{figure}[!t]
\centering
\subfigure[Before angle-aggregation.]{\includegraphics[width=0.47\linewidth]{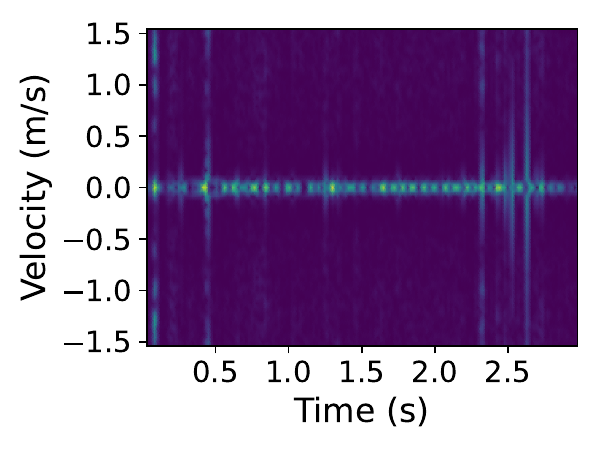}
    \label{fig:VT_before_fuse}}
\hfil
\subfigure[After angle-aggregation.]{\includegraphics[width=0.47\linewidth]{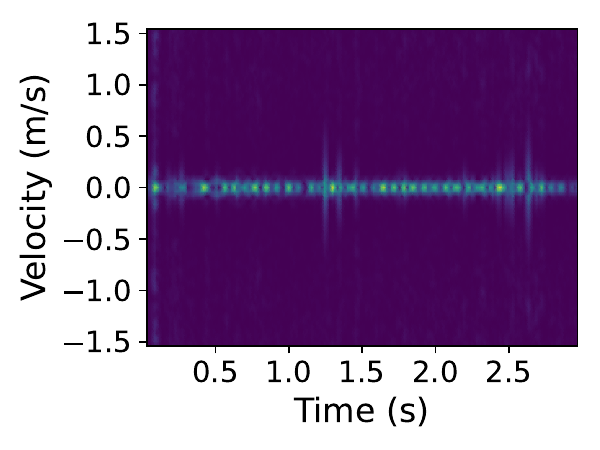}
    \label{fig:VT_after_fuse}}
\hfil
\vspace{-1em}
\caption{Comparison of time-frequency spectrum before and after multi-angle aggregation. The SNR improves from 4.05 dB to 10.44 dB.}
\label{fig:angle-aggregation}
\vspace{-1em}
\end{figure}

This aggregation reinforces the desired signal direction while suppressing off-angle noise and interference.
The effectiveness of this process is illustrated in Fig.~\ref{fig:angle-aggregation}. Compared to the unprocessed signal, the aggregated result exhibits significantly enhanced spectral clarity and phase consistency. Specifically, the signal-to-noise ratio (SNR) improves from 4.05 dB to 10.44 dB, indicating substantial suppression of noise components and improved tremor signal visibility.

\section{Secondary Reflection Enhancement}
\label{SEC:Secondary Relection Enhancement}
To address the challenge of obstructed LoS paths in real-world scenarios, we utilize secondary reflections, multipath signals that retain the hand's motion and phase characteristics, to assist the primary signal in reconstructing a complete and reliable tremor signal in this section.

\begin{figure}[!t]
\centering
\subfigure[Secondary reflection path.]{\includegraphics[width=0.4\linewidth]{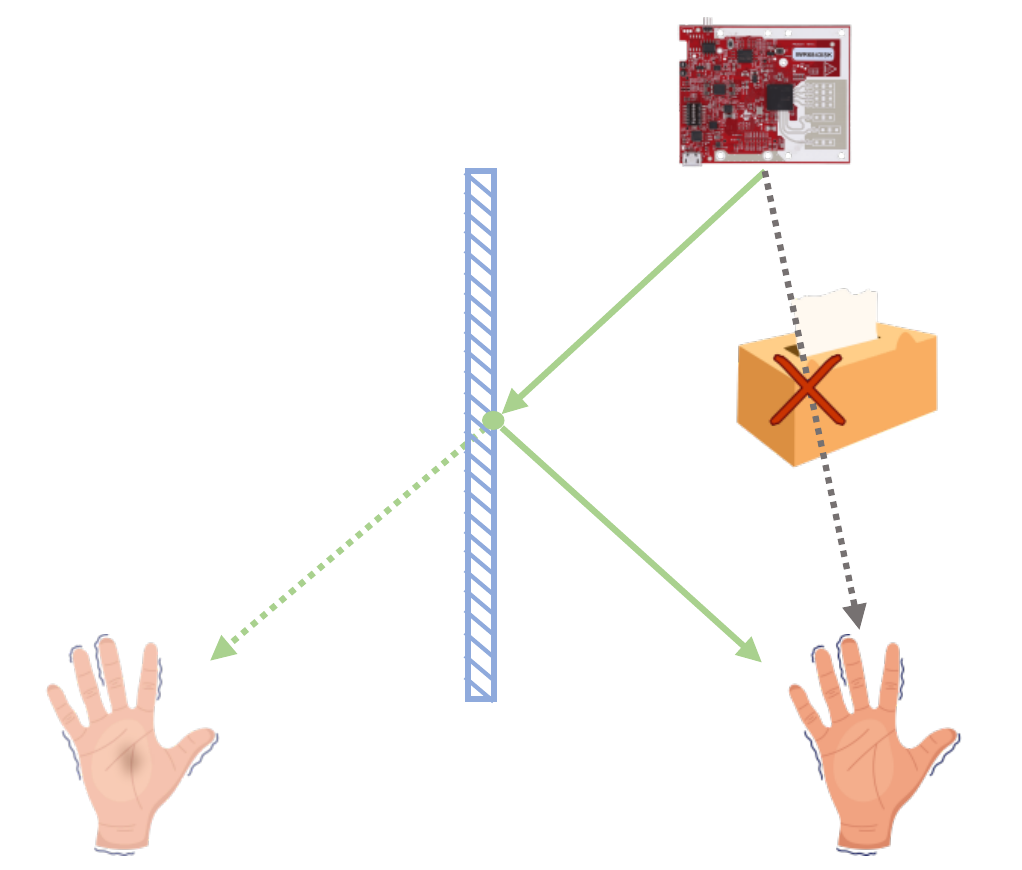}
    \label{fig:sec_illustration}}
\hfil
\subfigure[Beamforming result.]{\includegraphics[width=0.47\linewidth]{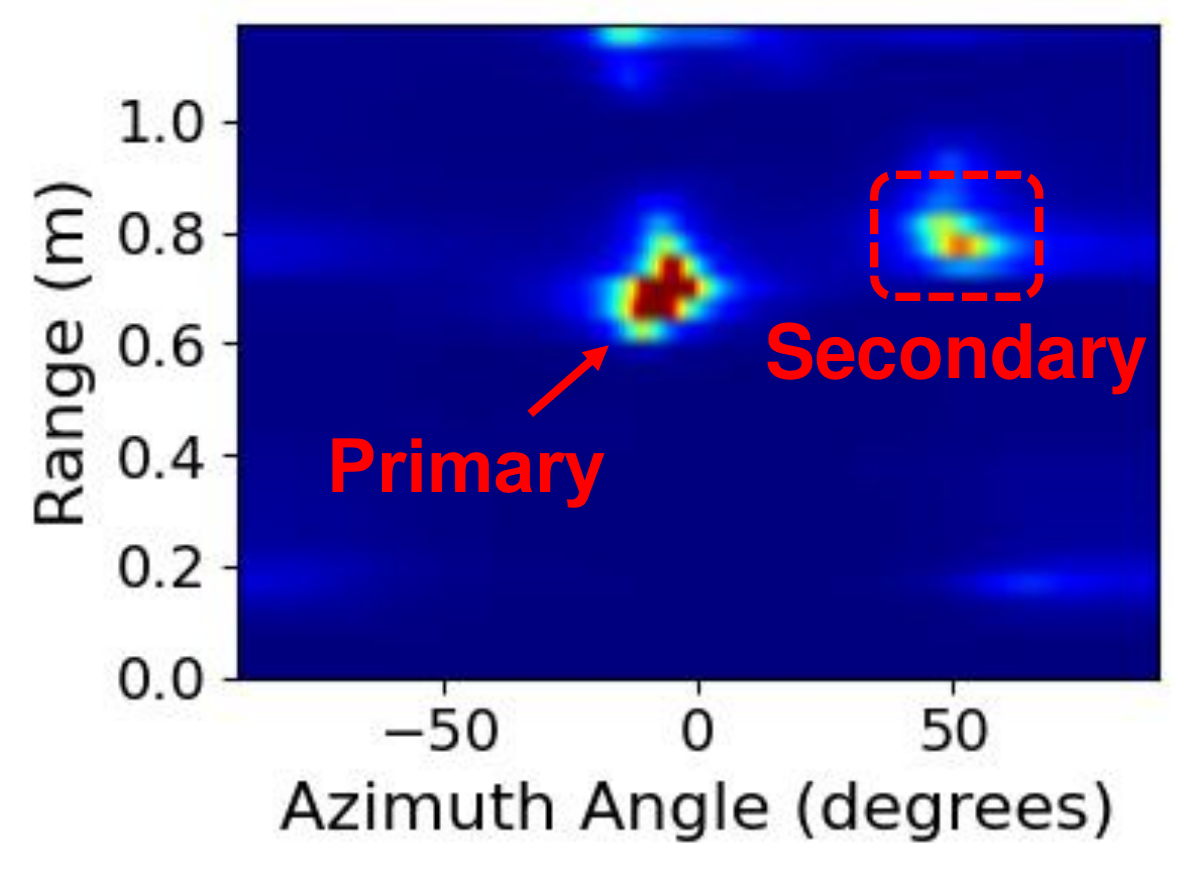}
    \label{fig:sec_beamforming}}
\vspace{-.5em}
\caption{Illustration and beamforming visualization of secondary reflection.}
\label{fig:second_reflection}
\vspace{-1em}
\end{figure}

\begin{figure}[!t]
\centering
\subfigure[Before secondary enhancement.]{\includegraphics[width=0.45\linewidth]{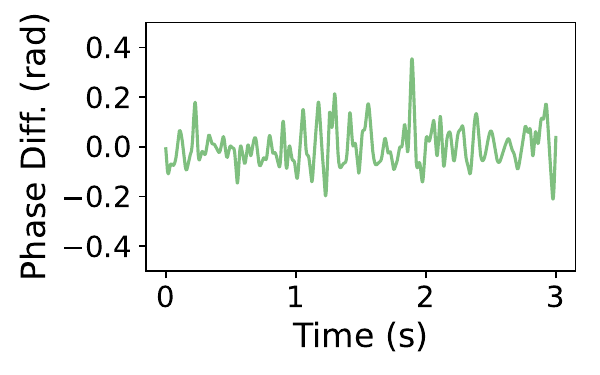}
    \label{fig:sec_phase}}
\hfil
\subfigure[After secondary enhancement.]{\includegraphics[width=0.45\linewidth]{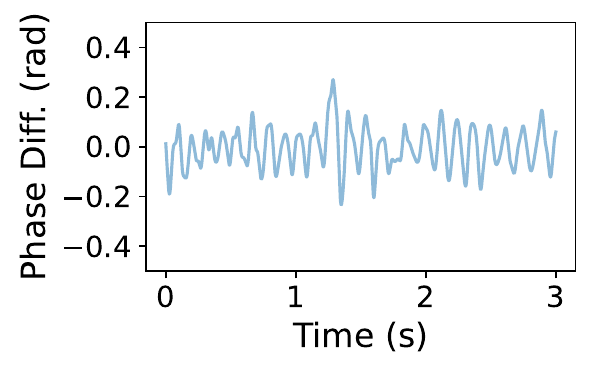}
    \label{fig:sec_phase_fused}}
\vspace{-.5em}
\caption{Effectiveness of secondary reflection enhancement in stabilizing hand tremor signals obtained from the primary path.}
\label{fig:secondary_enhancement}
\vspace{-1em}
\end{figure}

\subsection{Path Selection}
Although mmWave radar naturally generates multiple reflections due to environmental structures, only those corresponding to the hand are useful for tremor analysis. To identify valid secondary reflections from the same target, we evaluate their spectral and temporal similarity with a reference path. Specifically, we compute the Signal Variation-to-Noise Ratio (SVNR)~\cite{Gong2025SeRadar} and the phase trajectory correlation for each candidate path. SVNR is defined as
\begin{equation}
    \text{SVNR} = 10 \log_{10} \left( \frac{E_{\text{motion}}}{E_{\text{residual}}} \right)
\end{equation}
where $E_{\text{motion}}$ and $E_{\text{residual}}$ are the energy within and outside the tremor frequency band $[f_l, f_h]$. To obtain them, we apply FFT to the phase-difference signal $\phi(t)$ of each path, yielding a frequency-domain representation $\Phi(f)$.
Then the energy within and outside the band is defined as $E_{\text{motion}} = \frac{1}{N_s} \sum_{f \in [f_l, f_h]} |\Phi(f)|^2$ and $E_{\text{residual}} = \frac{1}{N_b} \sum_{f \notin [f_l, f_h]} |\Phi(f)|^2$, where $N_s$ and $N_b$ are the number of frequency bins inside and outside the tremor band, respectively.

To assess temporal similarity, we compute the Pearson correlation coefficient between the phase trajectory $\phi_i(t)$ of each candidate path and that of the reference $\phi_r(t)$ as

\begin{equation}
    \rho_i = \text{corr}(\phi_i(t), \phi_r(t))
\end{equation}

where $\phi_r(t)$ is typically the primary reflection; if unavailable, the strongest path is used as the reference.

Only candidate paths satisfying both $\text{SVNR}_i$ and $\rho_i$ above empirically determined thresholds are retained for fusion. This ensures that only tremor-relevant and phase-consistent reflections contribute to the final signal.

\subsection{Path Fusion}
To reconstruct a stable and high-fidelity tremor signal, we fuse the primary and selected secondary reflection paths through a three-step process: phase alignment, phase normalization, and phase-weighted fusion.

\paragraph{Phase Alignment}
Due to the differences in propagation delay and reflection geometry, the raw phase signals $\phi_i(t)$ of secondary paths may exhibit initial phase offsets. We align them to a common reference by computing the offset at an initial time $t_0$ and adjusting the full trajectory
\begin{equation}
    \hat{\phi}_i(t) = \phi_i(t) - \phi_i(t_0) + \phi_r(t_0)
\end{equation}
This step ensures that all paths begin with the same phase point at $t_0$, enabling alignment of temporal phase variations.

\paragraph{Phase Normalization}
Despite alignment, different paths may exhibit amplitude disparities caused by varying reflection angles and signal attenuation. To equalize their influence, we normalize the aligned phase $\hat{\phi}_i(t)$ within a sliding window $W$
\begin{equation}
    \tilde{\phi}_i(t) = \frac{\hat{\phi}_i(t) - \mu_i}{\sigma_i}, \quad t \in W
\end{equation}
where $\mu_i$ and $\sigma_i$ are the mean and standard deviation of $\hat{\phi}_i(t)$ in $W$. This normalization removes amplitude bias and ensures that all paths exhibit zero-mean, unit-variance behavior within each time window, thereby preventing any single path from dominating the fusion result due to scale differences alone.

\paragraph{Phase-Weighted Fusion}
Not all secondary paths contribute equally to signal quality as some may contain noise or weakly correlate with the tremor motion. To emphasize more informative paths, we introduce a quality-aware weighting mechanism based on SVNR. The fusion weight $w_i$ of each path is
\begin{equation}
    w_i = \frac{\text{SVNR}_i}{\sum_j \text{SVNR}_j}
\end{equation}
This normalized weighting scheme ensures that paths with high quality and strong tremor correlation contribute more. Finally, the enhanced hand tremor signal is obtained by a weighted summation of the normalized phases
\begin{equation}
    \phi_{\text{fused}}(t) = \sum_i w_i \cdot \tilde{\phi}_i(t)
\end{equation}

After the path phase alignment, normalization, and weighting, we successfully combine the selected reflection paths to enhance the overall hand tremor signal. The process of combining these paths results in a more stable and robust signal, as demonstrated by the phase difference comparison in Fig.~\ref{fig:secondary_enhancement}. Before the secondary reflection enhancement, the phase differences are unstable. However, after the enhancement process, the phase difference becomes clearer and more stable, leading to a better representation of the hand’s motion. It highlights the effectiveness of secondary reflection enhancement under partial occlusion conditions.

\textbf{\section{Tremor Monitoring}
\label{SEC:Tremor Monitoring}}
\begin{figure}[!t]
  \centering
  \includegraphics[width=1.0\linewidth]{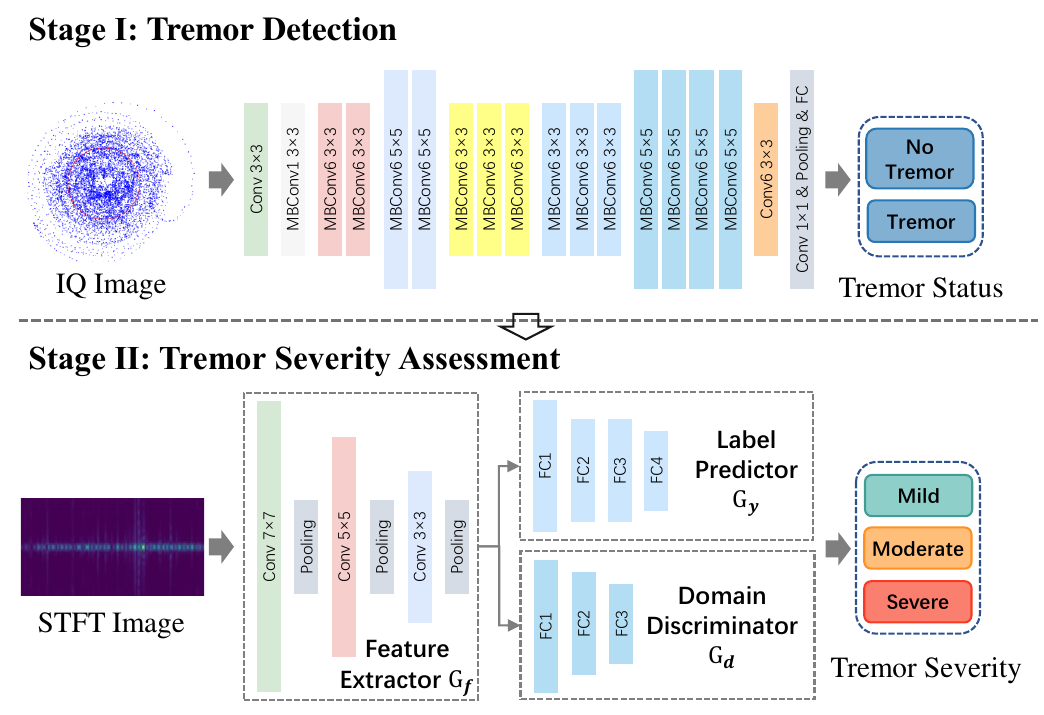}
  \vspace{-1.5em}
  \caption{Overview of the two-stage tremor monitoring framework.}
  \label{fig:two_stage_framework}
  \vspace{-1em}
\end{figure}

In this section, we propose a modular two-stage tremor monitoring framework that combines complementary time-domain and frequency-domain radar features to achieve high accuracy and robustness in tremor detection and severity assessment.

\subsection{Stage I: Tremor Detection.}
The first stage of the tremor monitoring framework aims to identify the presence of tremors using IQ images. The complex-valued baseband signal reflected from a moving target can be represented as $s(t) = I(t) + jQ(t)$, where $I(t)$ and $Q(t)$ denote the in-phase and quadrature components, respectively. By plotting these components over time in the complex plane, we obtain the IQ trajectory, which captures the micro-motion characteristics of the hand. For a non-stationary target, the distance and phase variations lead to time-varying IQ points whose distribution and path encode detailed motion patterns. These IQ trajectories are particularly sensitive to the nature of movement, whether the hand is static, gently moving, or experiencing high-frequency tremor. As a result, visual and statistical distinctions between IQ trajectories under different motion states become prominent and can be exploited for tremor detection tasks.

\begin{figure}[!t]
\centering
\vspace{-.5em}
\subfigure[Static state.]{\includegraphics[width=0.38\linewidth]{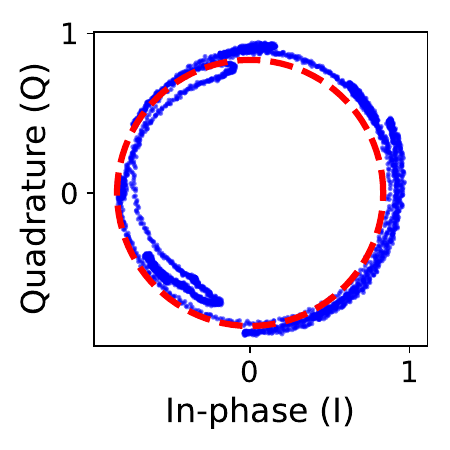}
    \label{fig:IQ_static}}
\hfil
\subfigure[Tremor state.]{\includegraphics[width=0.38\linewidth]{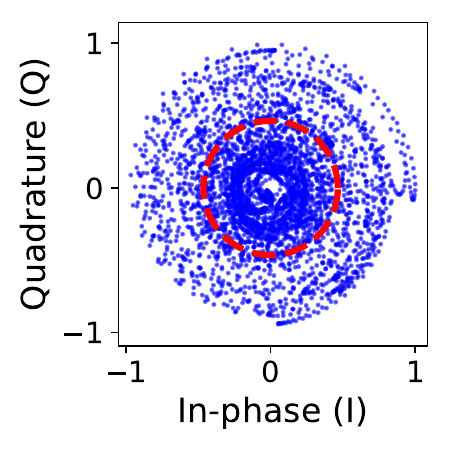}
    \label{fig:IQ_tremor}}
\vspace{-.5em}
\caption{Normalized IQ images of the hand in static and tremor states.}
\label{fig:IQ}
\vspace{-1em}
\end{figure}

In the static case, the hand remains largely motionless with only minimal perturbations arising from physiological sources such as respiration, heartbeat, or natural postural adjustments. These subtle motions result in the compact and continuous IQ trajectories, which are often quasi-periodic or drift slowly around a central point in the complex plane (as shown in Fig.\ref{fig:IQ_static}). In contrast, when the hand is affected by pathological or involuntary tremor, it undergoes rapid, rhythmic, and small-amplitude oscillations that are significantly higher in frequency than normal physiological motion. As a result, the IQ trajectory in the tremor state lacks the smoothness and coherence observed in the static case. Instead, it exhibits a noisy, diffused structure, often forming dense circular distributions or vortex-like patterns with poor continuity (as shown in Fig.\ref{fig:IQ_tremor}). These distinct visual and statistical patterns provide a discriminative basis for tremor detection.

To leverage these features, we formulate tremor detection as a binary classification task. Specifically, we employ a pre-trained EfficientNet fine-tuned on our dataset to classify IQ images into ``tremor'' or ``non-tremor''. The model captures both time-domain and frequency-domain features encoded in the IQ trajectory, making it particularly effective for differentiating between the subtle changes associated with tremor and non-tremor states. This stage serves as a crucial gatekeeper: if no tremor is detected, the system terminates further processing to reduce computational overhead and avoid false alarms.

\subsection{Stage II: Tremor Severity Assessment.}
\begin{figure}[!t]
\centering
\vspace{-.5em}
\subfigure[Subject A.]{\includegraphics[width=0.47\linewidth]{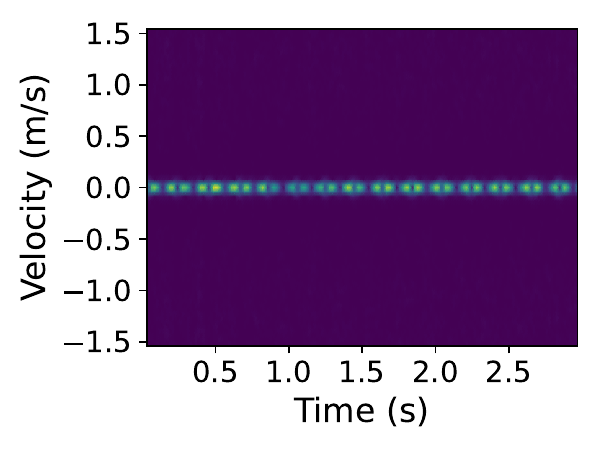}
    \label{fig:STFT2}}
\hfil
\subfigure[Subject B.]{\includegraphics[width=0.47\linewidth]{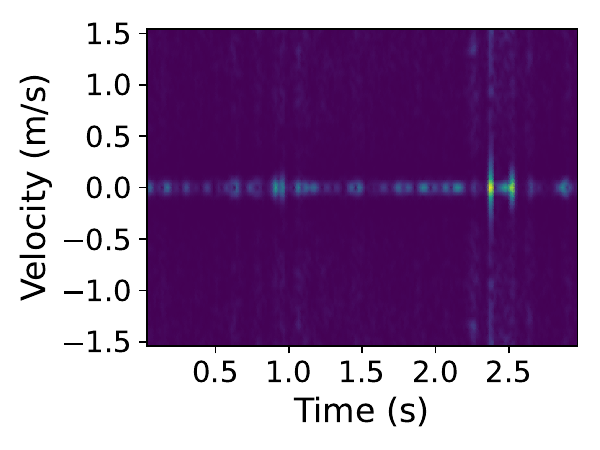}
    \label{fig:STFT1}}
\vspace{-.5em}
\caption{Micro-Doppler spectrograms of different subjects with mild tremor severity, showing user-specific variations in tremor frequency and amplitude.}
\label{fig:STFT}
\vspace{-1em}
\end{figure}
Once tremor is detected, the second stage assesses its severity by analyzing the micro-Doppler effect, which is able to capture fine-grained, periodic movements. Tremors, clinically defined as rhythmic, involuntary oscillatory movements, are characterized by distinct frequency and amplitude. These biomechanical properties are reflected in the micro-Doppler spectrogram in two key aspects: (1) \textit{Tremor Frequency}: Represented by symmetrical spectral lines centered around the zero-Doppler axis, with line density corresponding to the oscillation frequency. (2) \textit{Tremor Amplitude}: Influences the spectral bandwidth, with more severe tremors causing broader Doppler spreads due to increased movement velocity.

For tremor-positive samples, we first convert the phase difference signals into micro-Doppler spectrograms using STFT, which captures the frequency and amplitude characteristics of hand movements. These spectrograms are then fed into a CNN-based classifier trained to categorize tremor severity into three clinically relevant levels based on amplitude: (1) Mild (amplitude $<$ 2 cm); (2) Moderate (amplitude between 2 and 4 cm); (3) Severe (amplitude $>$ 4 cm)~\cite{elble2006tremor}.

While this classification framework is effective in principle, a major practical challenge arises from the substantial inter-subject variability in tremor characteristics. Specifically, different individuals may exhibit diverse tremor frequencies, amplitudes, and patterns, even within the same clinical tremor severity level. These variations are illustrated in Fig.~\ref{fig:STFT}, which shows spectrograms from two subjects labeled with the mild tremor but presenting distinct signal distributions. Such domain shifts can significantly degrade the generalization performance of conventional supervised learning models trained on limited, subject-specific data.

To address this, we adopt a domain adaptation strategy based on DANN, which enables the model to learn domain-invariant features that are robust to inter-subject differences. It consists of three components: a feature extractor $G_f$, a label predictor $G_y$, and a domain discriminator $G_d$. The overall training objective is

\begin{equation}
\mathcal{L} = \mathcal{L}_y(G_y(G_f(x_s)), y_s) - \lambda \cdot \mathcal{L}_d(G_d(G_f(x)), d)
\label{eq:dann_loss}
\end{equation}

where $\mathcal{L}_y$ is the cross-entropy loss on source domain samples $(x_s, y_s)$ and $\mathcal{L}_d$ is the binary cross-entropy loss of domain discriminator. The domain label $d \in \{0, 1\}$ indicates whether a sample belongs to the source or target domain, and $\lambda$ is a trade-off parameter that controls the adversarial learning strength. The gradient reversal layer (GRL) between $G_f$ and $G_d$ encourages the feature extractor to produce domain-invariant features, improving generalization across different domains.

\section{Prototype Implementation}
\label{SEC:Prototype Implementation}

\begin{figure}
    \centering
    \includegraphics[width=1.0\linewidth]{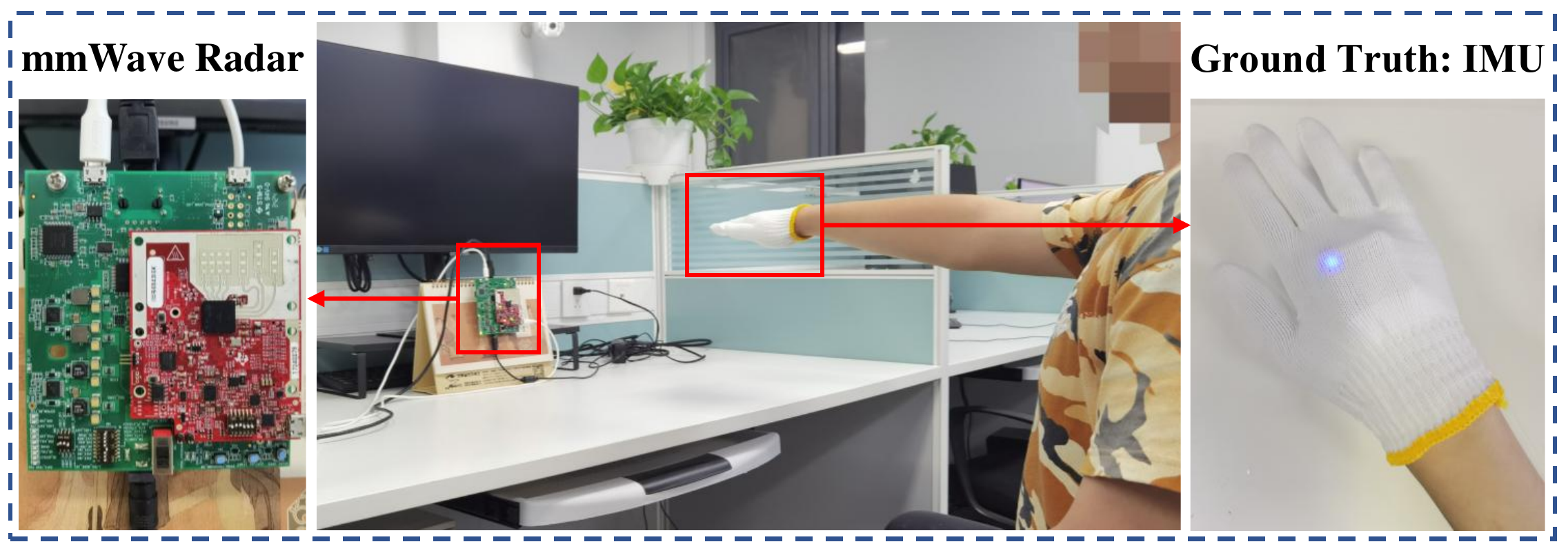}
    \vspace{-1.5em}
    \caption{An example of data collection. The COTS mmWave radar and IMU are used to capture experimental signals and ground truth.}
    \label{fig:implement}
    \vspace{-0.5em}
\end{figure}

\subsection{Data Collection and Annotation}
\begin{figure}
    \centering
    \includegraphics[width=1.0\linewidth]{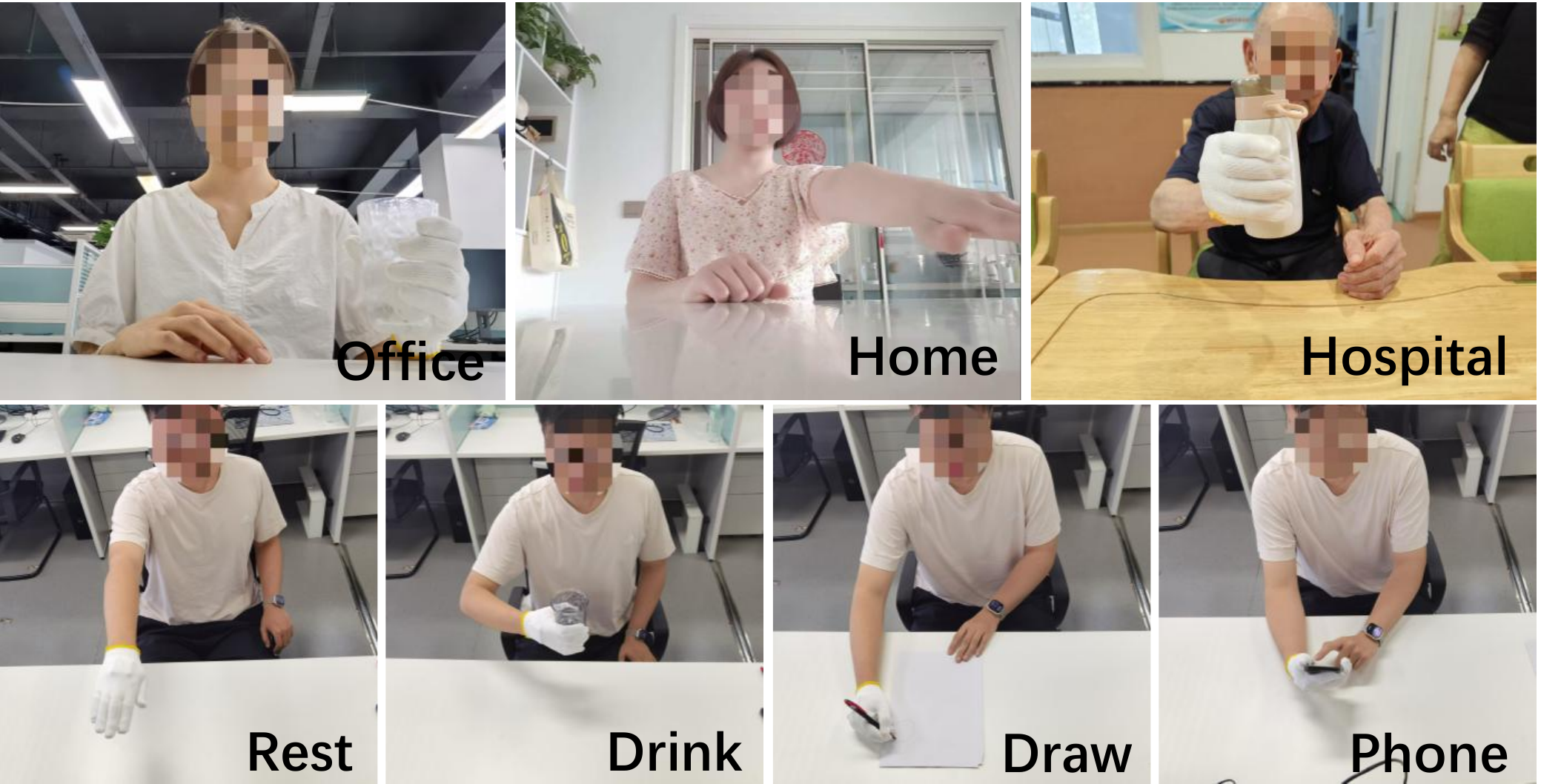}
    \vspace{-1em}
    \caption{Representative environments and common postures, covering typical usage scenarios and tremor types.}
    \vspace{-1em}
    \label{fig:envs}
\end{figure}

\begin{figure*}[!t]
	\begin{tabular}{cc}
		\begin{minipage}[t]{0.22\linewidth}
			\includegraphics[width = 1\linewidth]{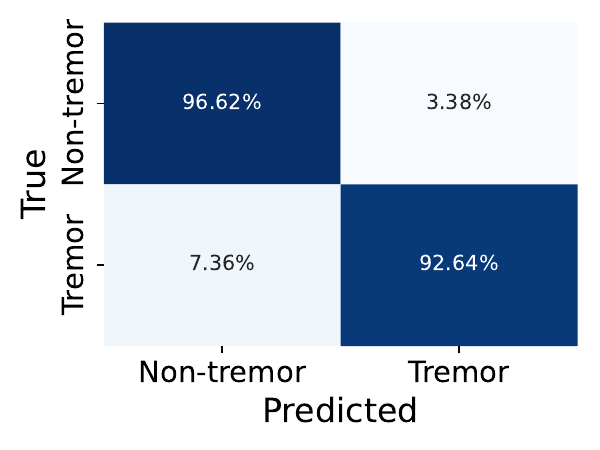}
                \vspace{-2em}
			\caption{Confusion matrix of tremor detection.}
			\label{fig:stage1_cm}
		\end{minipage}
		\quad
		\begin{minipage}[t]{0.22\linewidth}
			\includegraphics[width = 1\linewidth]{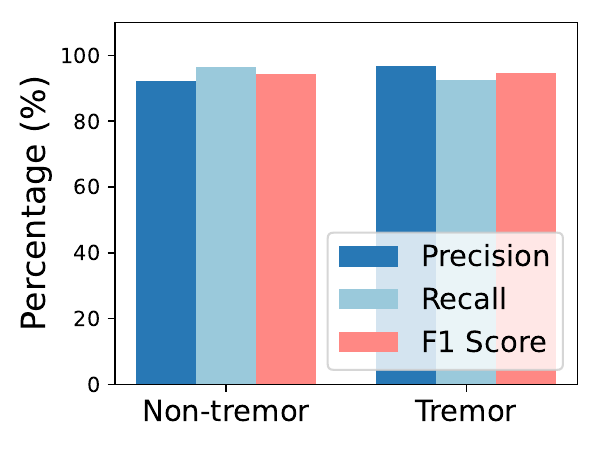}
                \vspace{-2em}
			\caption{Precision, recall and F1-score of tremor and non-tremor.}
			\label{fig:stage1_precision}
		\end{minipage}
        \quad
        \begin{minipage}[t]{0.22\linewidth}
			\includegraphics[width = 1\linewidth]{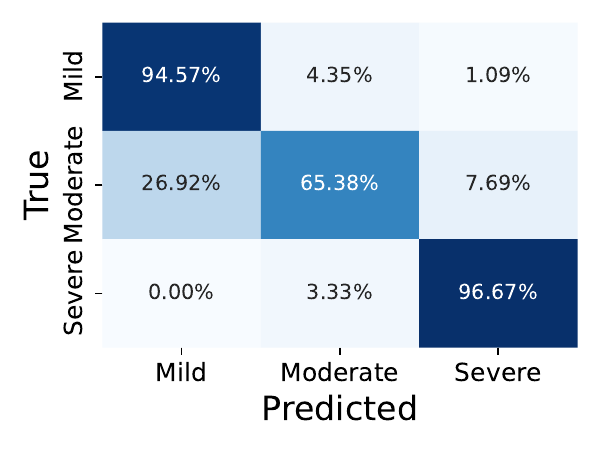}
                \vspace{-2em}
			\caption{Confusion matrix of tremor severity assessment.}
			\label{fig:tremor_severity_cm}
		\end{minipage}
		\quad
		\begin{minipage}[t]{0.22\linewidth}
			\includegraphics[width = 1\linewidth]{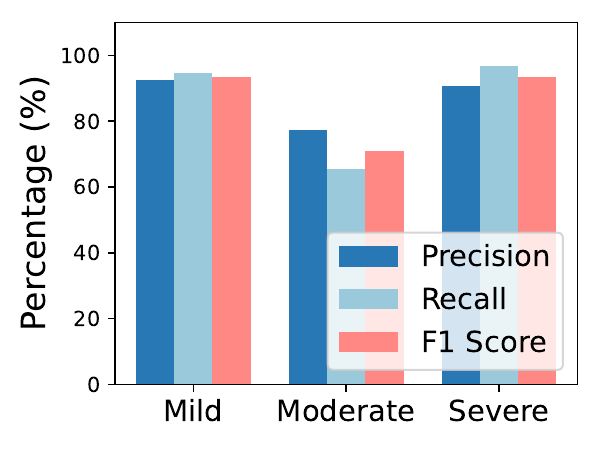}
                \vspace{-2em}
			\caption{Precision, recall and F1-score of each tremor level.}
			\label{fig:tremor_severity_precision}
		\end{minipage}
	\end{tabular}
    \vspace{-1em}
\end{figure*}
\begin{figure*}[!t]
	\begin{tabular}{cc}
		\begin{minipage}[t]{0.22\linewidth}
			\includegraphics[width = 1\linewidth]{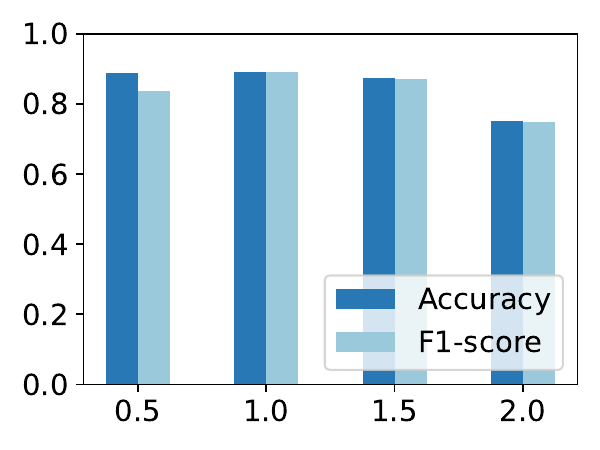}
                \vspace{-2em}
			\caption{Impact of distances (m).}
			\label{fig:eval_distances}
		\end{minipage}
        \quad
        \begin{minipage}[t]{0.22\linewidth}
			\includegraphics[width = 1\linewidth]{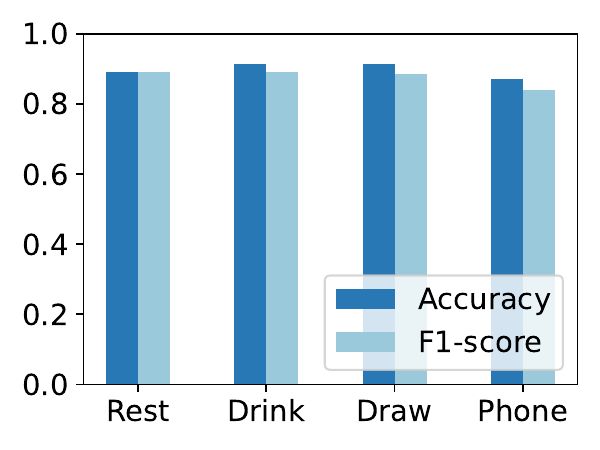}
                \vspace{-2em}
			\caption{Impact of postures.}
			\label{fig:eval_postures}
		\end{minipage}
        \quad
        \begin{minipage}[t]{0.22\linewidth}
			\includegraphics[width = 1\linewidth]{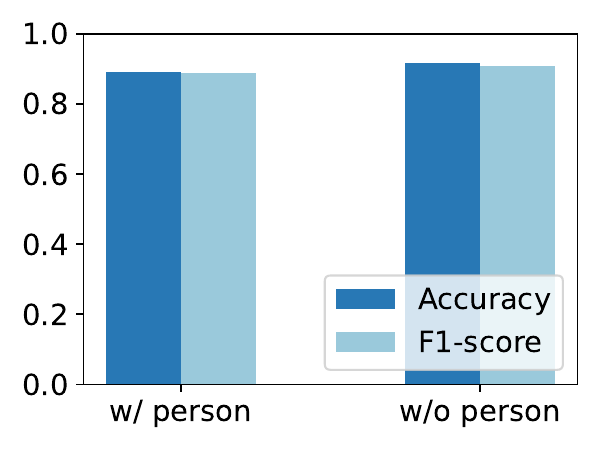}
                \vspace{-2em}
			\caption{Impact of interferences.}
			\label{fig:eval_interferences}
		\end{minipage}
		\quad
        \begin{minipage}[t]{0.22\linewidth}
			\includegraphics[width = 1\linewidth]{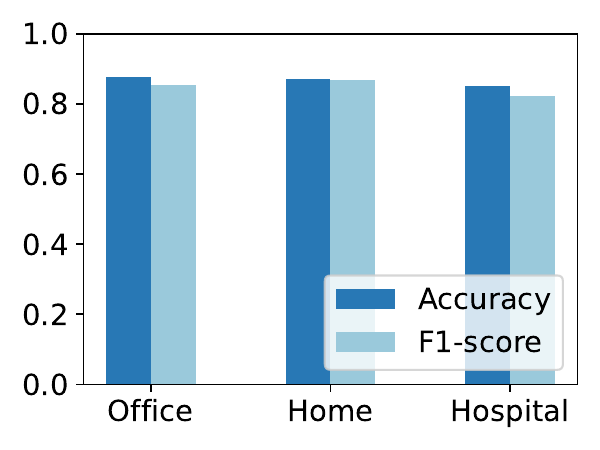}
                \vspace{-2em}
			\caption{Impact of environments.}
			\label{fig:eval_envs}
		\end{minipage}
	\end{tabular}
    \vspace{-1em}
\end{figure*}

\subsubsection{Participants Recruitment}
We recruit 30 volunteers from local universities, hospitals, and nursing homes, including 5 clinically diagnosed tremor patients and 25 healthy individuals. The participants range in age from 22 to 91 years (mean: 52 $\pm$ 18). IMU measurements reveal a wide distribution of tremor amplitudes, ranging from 0.07 to 16.04 cm (mean: 2.09 $\pm$ 2.50 cm). All procedures are reviewed and approved by the Institutional Review Board (IRB) of the hosting institution.

\subsubsection{Ground Truth Acquisition}
Ground truth is obtained using an LPMS-B2 IMU affixed to the back of a wearable glove on the dominant hand (Fig.~\ref{fig:implement}). The sample rate of IMU is 50 Hz. Following \cite{hu2025mmTremor}, segments with an amplitude greater than 5 mm are labeled tremor. Tremor severity level is determined based on the average amplitude of the displacement.

\subsection{System Implementation}
\subsubsection{Sensing Platform}
Our sensing platform is TI IWR6843 mmWave radar sensor (Fig.~\ref{fig:implement}), operating at 60 GHz with a 4 GHz bandwidth.
Each frame consists of 64 chirps, with 256 ADC samples per chirp at a 5 Msps sampling rate. Frames are transmitted at 25 Hz.

\subsubsection{Model Training and Parameter Setting}
Both the IQ- and STFT-based image models are optimized using Adam with a fixed learning rate of $10^{-3}$. For tremor detection, EfficientNet-B0 is trained for 20 epochs with a batch size of 16. For tremor severity assessment, we train the DANN-based network using a batch size of 8 for 100 epochs.

\subsection{Experiment Setup}
During the experiment, each participant was seated 1 m in front of the radar device. The healthy subjects were instructed to simulate tremor-like hand movements. For each of them, we collected a total of 24 samples: 12 samples under static conditions and 12 samples under simulated tremor conditions. For patients with tremors, we collected 24 natural tremor samples without any intervention, capturing their spontaneous hand movement patterns. Each sample lasts 3 seconds.

\section{Evaluation}
\label{Sec:Evaluation}
In this section, we conduct extensive experiments to evaluate the performance of \name under various conditions.

\subsection{Evaluation Metrics}
To evaluate the proposed system, we use standard classification metrics including accuracy and class-imbalance-aware metrics such as weighted precision, recall, and F1-score. In addition, confusion matrices are presented to visualize inter-class performance and highlight misclassification patterns. All reported results are averaged over 10-fold cross-validation to ensure robustness and reliability.

\subsection{Micro-benchmark Studies}
Most existing work focuses primarily on detecting the existence of tremors. To maintain comparability and completeness, we also evaluate our system on this binary classification task. As shown in Fig.~\ref{fig:stage1_cm}, the confusion matrix illustrates that the model achieves high discriminative performance for both tremor and non-tremor classes. On average, the system achieves an accuracy of 94.51\%, outperforming existing works (TABLE~\ref{tab:comparison}) such as the IMU-based method~\cite{timmermans2025generalizable} with 61\% sensitivity and the mmWave radar-based method~\cite{hu2025mmTremor} with 89.7\% detection accuracy.
To provide a more detailed breakdown, Fig.~\ref{fig:stage1_precision} presents the precision, recall, and F1-score for each class. All three metrics exceed 90\%, confirming the balanced performance of the model. These results highlight the model’s robustness and reliability in binary tremor detection tasks. It achieves both low false-positive and false-negative rates, which is essential for clinical screening scenarios.

\begin{figure}[!t]
\centering
\vspace{-.5em}
\subfigure[$\alpha = 0$]{\includegraphics[width=0.42\linewidth]{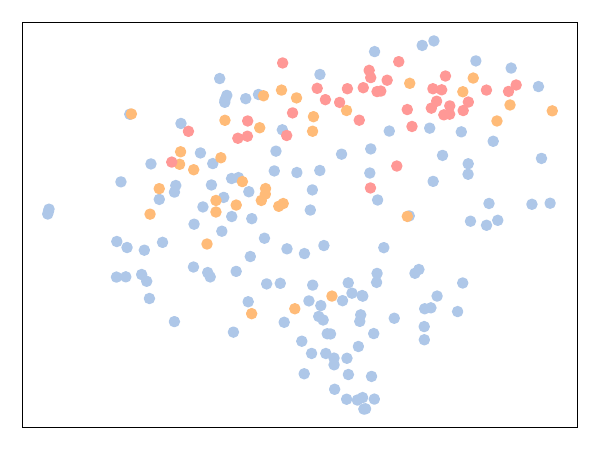}
    \label{fig:tsne_before}}
\hfil
\subfigure[$\alpha = 1$]{\includegraphics[width=0.42\linewidth]{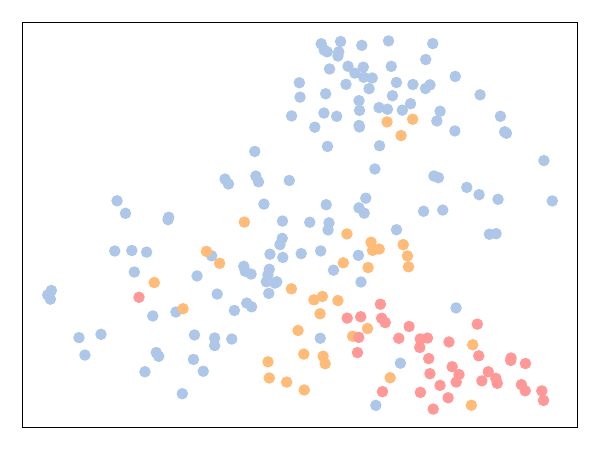}
    \label{fig:tsne_after}}
\vspace{-.5em}
\caption{t-SNE visualization of the feature embeddings w/o ($\alpha = 0$) and w/ ($\alpha = 1$) domain adaptation. The blue, yellow, and red dots represent the ``Mild'', ``Moderate'', and ``Severe'' tremor severity levels, respectively.}
\label{fig:tsne}
\vspace{-1em}
\end{figure}

In addition, we conduct a transferability analysis by visualizing the t-SNE map of feature embeddings with and without domain adaptation. In Fig.~\ref{fig:tsne_before}, there is a considerable overlap in the t-SNE map, indicating limited generalization. However, with the domain adaptation scheme, as shown in Fig.~\ref{fig:tsne_after}, the classes are more distinctly separated in the feature space, confirming an improvement in the model's ability to generalize.

\subsection{Overall Performance}
In real-world clinical scenarios, assessing the severity of tremor is equally important for diagnosis and treatment planning. We evaluate the model's capability in the classification of tremor severity levels. On the test set, the model achieves an overall accuracy of 89.13\%, a precision of 89.75\%, a recall of 89.13\%, and an F1 score of 88.72\%. These results demonstrate that the proposed system can effectively differentiate between various tremor severity levels. Compared with existing methods, our system offers a more comprehensive solution. While camera-based approaches can capture tremor with a precision of 2.6 mm (TABLE~\ref{tab:comparison}), they raise privacy concerns. In contrast, privacy-preserving solutions like mmTremor~\cite{hu2025mmTremor} do not support severity assessment. Our model bridges this gap by enabling accurate and privacy-friendly tremor severity assessment.
As shown in Fig.~\ref{fig:tremor_severity_cm}, the model performs well in identifying mild and severe tremors, with correct classification rates of 94.57\% and 96.67\%, respectively. However, moderate tremors remain challenging, with only 65.38\% correctly classified. A notable portion of moderate samples are misclassified as mild (26.92\%) or severe (7.69\%). Mild and severe tremors are easily misclassified to the adjacent class. The probable reasons are further discussed in Sec.~\ref{SEC:Discussion}.

\subsection{Robustness and Generalization}
\begin{figure}[!t]
	\begin{tabular}{cc}
		\begin{minipage}[t]{0.45\linewidth}
			\includegraphics[width = 1\linewidth]{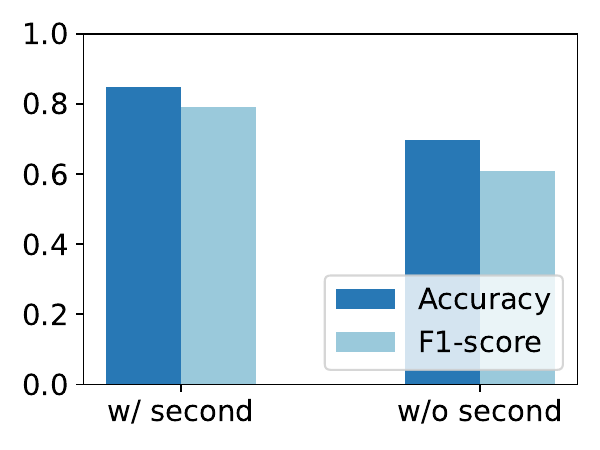}
                \vspace{-2em}
			\caption{Impact of secondary reflection enhancement.}
			\label{fig:eval_second}
		\end{minipage}
		\quad
		\begin{minipage}[t]{0.45\linewidth}
			\includegraphics[width = 1\linewidth]{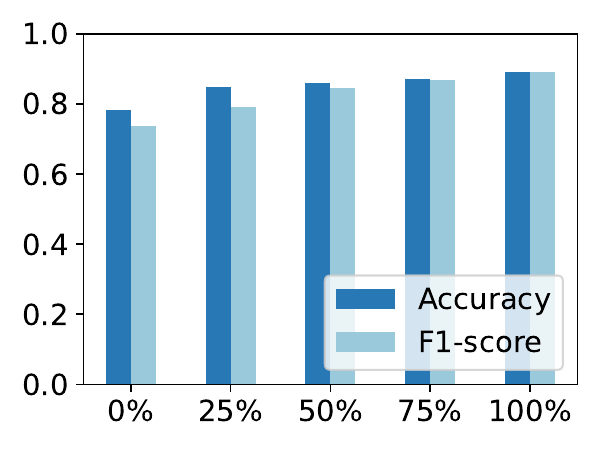}
                \vspace{-2em}
			\caption{Impact of domain adaptation rate.}
			\label{fig:eval_adaptation}
		\end{minipage}
	\end{tabular}
    \vspace{-1em}
\end{figure}

\subsubsection{Impact of Distance}
The distance between the sensor and the user can vary depending on the use scenario. To identify the optimal deployment range, we evaluate system performance at distances from 0.5 to 2.0 m.
As shown in Fig.~\ref{fig:eval_distances}, the system maintains a high accuracy and F1-score above 0.85 within a 1.5 m range, ensuring reliable performance for desktop or seated activities. At 2.0 m, performance slightly drops due to signal attenuation and multipath interference. These findings offer practical guidance for deployment, suggesting that the system should ideally operate within 1.5 meters for optimal reliability in real-world environments.

\subsubsection{Impact of Postures}
Tremors can occur during a patient's daily activities, which may influence the tremor's intensity and frequency.
We evaluate the impact of different postures and object interactions, including resting, drinking, drawing, and using a phone (Fig.~\ref{fig:envs}).
The results are shown in Fig.~\ref{fig:eval_postures}.
Despite changes in hand posture and micro-movement patterns, the system performs well for tremor severity assessment, indicating good generalization of natural hand behaviors.

\subsubsection{Impact of Interference}
In practical use, it is inevitable that there will be other individuals moving around the patient. Therefore, we test the system performance with or without pedestrian interference.
The experimental results (Fig.~\ref{fig:eval_interferences}) indicate that the presence of nearby pedestrians has a negligible impact on the system's accuracy. The tremor assessment module maintains consistent performance, with fewer than 3\% changes in accuracy between trials. This can be attributed to the fact that the system is designed to first isolate the hand region from surrounding interference and subsequently enhance the hand-specific signals, thereby suppressing unrelated background motion such as pedestrian movement.

\subsubsection{Impact of Environment}
The system is expect to operate reliably across diverse real-world environments. To evaluate its adaptability, we conduct experiments in three representative settings: an office, a home, and a clinical hospital, as shown in Fig.\ref{fig:envs}. The results presented in Fig.\ref{fig:eval_envs} demonstrate that the system consistently achieves accuracy and F1-scores above 0.85 in all environments, indicating strong robustness against variations in background clutter, ambient noise, and environmental dynamics.

\subsection{Ablation Study}

To evaluate the contribution of our designed system component, we conduct ablation studies on the secondary reflection enhancement part and the domain adaptation part.
Fig.~\ref{fig:eval_second} presents the performance comparison with and without secondary reflection enhancement. Incorporating this mechanism yields substantial gains in both accuracy and F1-score, demonstrating its effectiveness in improving signal robustness under NLOS conditions.
We also investigate the effect of varying domain adaptation rates (0\%, 25\%, 50\%, 75\%, and 100\%), as shown in Fig.~\ref{fig:eval_adaptation}. Performance improves progressively with higher adaptation rates. These results demonstrate that even a moderate amount of domain-specific data enables the model to better adapt to distribution shifts.

\section{Discussion}
\label{SEC:Discussion}

\begin{figure}[!tb]
	\begin{tabular}{cc}
		\begin{minipage}[t]{0.45\linewidth}
			\includegraphics[width = 1\linewidth]{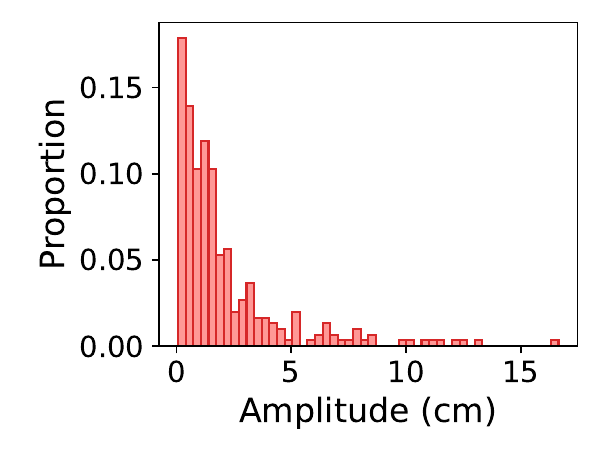}
                \vspace{-2em}
			\caption{Distribution of tremor amplitudes across all samples.}
			\label{fig:amp_dis}
		\end{minipage}
        \quad
		\begin{minipage}[t]{0.45\linewidth}
			\includegraphics[width = 1\linewidth]{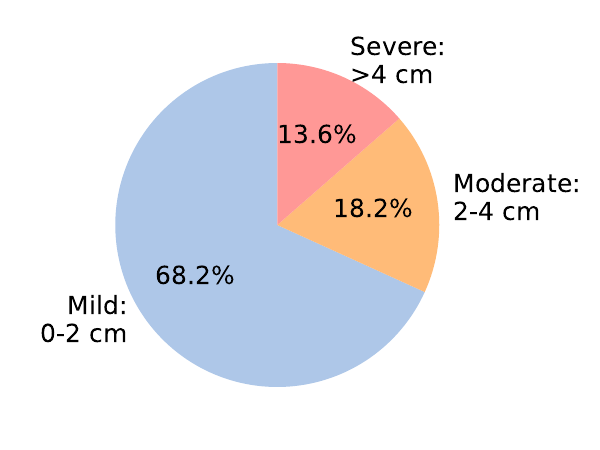}
                \vspace{-2em}
			\caption{Distribution of tremor amplitudes by severity levels.}
			\label{fig:freq_dis}
		\end{minipage}

	\end{tabular}
    \vspace{-1em}
\end{figure}

\begin{table}[!t]
\centering
\renewcommand{\arraystretch}{1.3}
\caption{Statistics on Tremor Amplitudes of Severity Levels}
\label{tab:statistics}
\begin{tabular}{|c|c|c|c|}
\hline
\textbf{} & \textbf{Mild (0-2cm)} & \textbf{Moderate (2-4cm)} & \textbf{Severe (\textgreater{}4cm)} \\ \hline
Avg.      & 0.86                  & 2.80                      & 7.31                                \\ \hline
Min.      & 0.07                  & 2.01                      & 4.04                                \\ \hline
Max.      & 1.98                  & 3.91                      & 16.64                               \\ \hline
Std.      & 0.30                  & 0.34                      & 8.67                                \\ \hline
\end{tabular}
\vspace{-1.5em}
\end{table}

\begin{table*}[!t]
\renewcommand{\arraystretch}{1.3}
\centering
\begin{threeparttable}
\caption{Comparison of Tremor Monitoring Methods}
\label{tab:comparison}
\begin{tabular}{c|cccccccc}
\Xhline{1.1pt}
                                  &                                        &                                                                                          &                                        & \multicolumn{2}{c}{\textbf{Tremor}}                                              &                                                                                           \\
\multirow{-2}{*}{\textbf{Method}} & \multirow{-2}{*}{\textbf{Sensor Type}} & \multirow{-2}{*}{\textbf{\begin{tabular}[c]{@{}c@{}}Privacy-\\ protecting\end{tabular}}} & \multirow{-2}{*}{\textbf{Non-contact}} & \multicolumn{1}{l}{\textbf{Detection}} & \multicolumn{1}{l}{\textbf{Assessment}} & \multirow{-2}{*}{\textbf{\begin{tabular}[c]{@{}c@{}}Works under\\ Occlusion\end{tabular}}} & \multirow{-2}{*}{\textbf{Accuracy\tnote{1}}} & \multirow{-2}{*}{\textbf{Data\tnote{2}}} \\ \hline
\cite{timmermans2025generalizable}           &   IMU                 &   \usym{2713}   &   \usym{2717}   &   \usym{2713}   &   \usym{2717}   &   \usym{2713} & Se: 61\% Sp: 97\% & P \\
\cite{friedrich2024validation}           &   camera          &   \usym{2717}   &   \usym{2713}   &   \usym{2713}   &   \usym{2713}   &   \usym{2717} & 2.6 mm & P \\ \hline
\cite{blumrosen2011noncontact}           &   UWB                 &   \usym{2713}   &   \usym{2713}   &   \usym{2713}   &   \usym{2713}   &   \usym{2717} & 1 mm & M \\ \hline
mmTremor\cite{hu2025mmTremor}           &   mmWave+depth camera &   \usym{2713}   &   \usym{2713}   &   \usym{2713}   &   \usym{2717}   &   \usym{2717} & D: 89.7\% & H+P \\
\textbf{\name} & mmWave        & \usym{2713}     & \usym{2713}     & \usym{2713}     & \usym{2713}     & \usym{2713} & D: 94.5\% A: 89.1\% & H+P \\ \Xhline{1.1pt}
\end{tabular}
\begin{tablenotes}
        \footnotesize
        \item[1] Se and Sp refer to the sensitivity and specificity of tremor detection. D and A refer to the accuracy of tremor detection (D) and tremor severity assessment (A), respectively.
        \item[2] M: Data collected using mechanical simulation. H: Data from healthy subjects simulating tremor. P: Data from clinically diagnosed patients' real tremor.
\end{tablenotes}
\end{threeparttable}
\vspace{-1em}
\end{table*}

To further investigate the reasons for the differences in accuracy at different levels, we analyze the amplitude statistics of each tremor level. Fig.~\ref{fig:freq_dis} reveals that mild tremor accounts for the largest proportion among all the data. So the model has sufficient data to learn features of mild tremor, resulting in the highest accuracy. Although the size of the moderate dataset is larger than the severe one, the accuracy of the severe tremor is higher in comparison. It is influenced by boundary cases between tremor levels. For instance, the minimum and maximum amplitudes of moderate tremor are 2.01 cm and 3.91 cm (as shown in TABLE.~\ref{tab:statistics}), which are close to the upper threshold of the mild level and the lower threshold of the severe level, leading to easy misclassifications. Furthermore, the variance of severe tremor is much greater than that of moderate tremor, making the features more distinctive. In clinical practice, boundary cases are also challenging for doctors to differentiate, as they typically rely on visual inspection, which is subjective and imprecise. Therefore, it is acceptable for our system to sometimes misclassify boundary cases into adjacent levels. In future work, we plan to further explore and analyze the impact of these boundary cases on the system’s performance and recruit a larger and more diverse patient cohort to ensure balanced representation across all tremor severity levels, especially in boundary cases. This will facilitate more accurate modeling of transitional patterns and improve the system’s performance in real-world clinical scenarios.

\section{Related Work}
\label{SEC:Related Work}

Currently, research on tremor monitoring in daily life primarily falls into three categories: wearable device-based methods, vision-based methods, and RF-based methods.

\subsection{Wearable Device-based Tremor Monitoring Methods}
Wearable devices have been widely used for tremor monitoring, as they provide direct motion measurements through accelerometers, gyroscopes, and electromyography (EMG) sensors \cite{rovini2017wearable, sigcha2023deep}. However, their practicality in daily life is limited by user discomfort, the need for regular wearing and maintenance, and the risk of being forgotten, particularly among elderly individuals. Additionally, wearable solutions inherently rely on user compliance, which can impact long-term usability.

\subsection{Vision-based Tremor Monitoring Methods}
Vision-based methods utilize cameras and computer vision algorithms to monitor tremor behavior in a non-contact manner. For example, Friedrich et al.~\cite{friedrich2024validation} employed pose tracking algorithms to analyze hand tremors from clinical video recordings. \cite{liu2023vision } estimates MDS-UPDRS tremor scores using optical flow and temporal difference features. \cite{alper2020pose} uses the Kinect 2 depth camera to achieve non-contact measurement of limb tremors. While vision-based systems offer real-time monitoring without physical attachments, they face significant challenges in privacy protection, as continuous video recording can intrude on personal spaces such as homes or healthcare environments. Additionally, their performance is highly sensitive to environmental factors like lighting and occlusions, which limit robustness and reliability in complex real-world settings.

\subsection{RF-based Tremor Monitoring Methods}
RF-based methods, including ultra-wideband (UWB) systems~\cite{blumrosen2011noncontact} and mmWave radar~\cite{mejdani2024radar, hu2025mmTremor}, have emerged as privacy-preserving, non-contact alternatives. Blumrosen et al.~\cite{blumrosen2011noncontact} propose a UWB-based system that detects tremor by analyzing periodicity in signal reflections. \cite{mejdani2024radar} utilizes FMCW radar and a neural network to estimate tremor frequency in clinical-like tasks. \cite{hu2025mmTremor} enables robust tremor detection during daily activities by combining a depth camera and a mmWave radar. These technologies detect subtle body movements using radio waves, enabling continuous tremor monitoring without requiring users to wear devices. Unlike vision-based systems, RF sensing is robust to lighting conditions and offers better occlusion tolerance in some cases. Nevertheless, existing RF-based solutions mostly focus on tremor detection rather than quantitative severity assessment, and their ability to operate under occlusion remains limited.

Our proposed \name is an mmWave radar-based system specifically designed for daily-life tremor monitoring. It offers non-contact, privacy-friendly operation, supports simultaneous tremor detection and severity assessment, and remains effective even under occlusion. Table~\ref{tab:comparison} highlights how \name advances tremor monitoring technologies from various aspects.

\section{Conclusion}
\label{SEC:Conclusion}

In this work, we introduce \name, a non-contact, privacy-friendly tremor monitoring system using mmWave radar, designed for tremor detection and severity assessment. We propose a series of tailored algorithms, including adaptive range beamforming, secondary reflection signal enhancement, and unsupervised domain adaptation to improve generalization. Experimental results from various real-world scenarios show that \name achieves high accuracy in tremor detection and severity assessment, making it a promising tool for home-based monitoring and early intervention in patients with neurological disorders. However, limitations remain in detection range and in monitoring patients during motion, which we aim to address in future work.

\section*{Acknowledgment}

The research is partially supported by China National Natural Science Foundation with No. 62132018, 62231015, ``Pioneer'' and ``Leading Goose'' R\&D Program of Zhejiang, 2023C01029 and 2023C01143, the PFP of CPSF with No. GZC20251058, and AJSP of CPSF with No. 2025T022AH.

\bibliographystyle{plain}
\bibliography{reference}

\end{document}